# An RF Timer of Electrons and Photons with the Potential to reach Picosecond Precision

Amur Margaryan[*], Vanik Kakoyan, Simon Zhamkochyan, Sergey Abrahamyan, Hayk Elbakyan, Samvel Mayilyan, Henrik Vardanyan, Hamlet Zohrabyan, Lekdar Gevorgian, Robert Ayvazyan, Artashes Papyan, Garnik Ayvazyan

A.I. Alikhanyan National Science Laboratory (Yerevan Physics Institute) Foundation, Yerevan, Armenia

Bagrat Grigoryan

CANDLE Synchrotron Research Institute, Yerevan, Armenia

John Annand, Kenneth Livingston, Rachel Montgomery

School of Physics & Astronomy, University of Glasgow, G12 8QQ Scotland, UK

Patrick Achenbach, Josef Pochodzalla

Institut für Kernphysik, Johannes Gutenberg-Universität Mainz, Mainz, Germany

Dimiter L. Balabanski

Extreme Light Infrastructure- Nuclear Physics (ELI-NP), Bucharest-Magurele, Romania

Satoshi N. Nakamura

Department of Physics, Graduate School of Science, the University of Tokyo, Tokyo, Japan

*corresponding author: mat@mail.yerphi.am

**Abstract.** This paper describes a new radio frequency timer of keV energy electrons. It is based on a helical deflector, which performs circular or elliptical sweeps of keV electrons, by means of 500 MHz radio frequency field. By converting a time distribution of incident electrons to a hit position distribution on a circle or ellipse, this device achieves extremely precise timing. Streak Cameras, based on similar principles, routinely operate in the ps and sub-ps time domain, but have substantial dead time associated with the readout system. Here, we report a new type of RF timing technique, where the position sensor, consisting of microchannel plates and a delay-line anode, produces ~ns duration pulses with small dead time. Measurements made with sub-ps duration laser pulses, synchronized to the radio frequency power, produced a timing resolution of ~10 ps. This ultra-high precision technique has potential applications in a large variety of scientific devices, and in all cases, electrons are timed and detected simultaneously in the same device.

## 1. Introduction and Overview

Precise measurements of time intervals between two or more physical events or between a physical event and a clock are needed in many applications in science and industry. In a typical timing technique, the time interval is measured between the leading edges of two electronic pulses applied to the start and stop inputs of a time-interval meter. A typical circuit might measure the difference in arrival time of two photons. The detectors, e.g. vacuum or Si photomultipliers, produce ~ns rise time pulses, with constant-fraction discriminators (CFD) providing sub-ns, time-pick-off precision for the logic pulses fed to time-to-digital converters (TDC).

Scientific applications of high-precision time measurement techniques occur in physics, astronomy, chemistry, biomedical imaging and material science. Technological and industrial applications include [1,2] laser ranging, depth imaging, dynamic testing of integrated circuits and high-speed optical components for data storage and fiber optic telecommunication. High-precision time difference measurements at high repetition rates, e.g. between two photon pulses or between clock and photon pulses, cannot currently achieve picosecond resolution. Modern digital circuits can operate at clock speeds of some tens of GHz, but they are not yet fast enough to directly count ps intervals. Nevertheless, TDCs implemented using high-speed digital circuits, are standard equipment for time measurement. There are many variants of TDC (see e.g. [3–5]). The simplest implementation of a TDC is a digital counter running at the speed of a fast crystal locked clock. Fast semiconductor technologies can in principle operate at clock rates as high as 40 GHz, thereby providing a time resolution of 25 ps. Time resolution may be improved beyond the clock period with interweaved electronic systems [2], but the dead time will always be limited. To achieve a resolution of 1 ps one must currently accept a TDC dead time of typically 80 ns, while 25 ps resolution entails about 25 ns dead time [6,7]. However, there are other practical limitations to achievable timing performance. A TDC with 1 ps resolution and zero dead time is of little use if the photon detector imposes a much larger time dispersion and dead time. At present, the detection of optical signals, down to the single-photon level, may be carried out with Avalanche Photodiodes (APD), vacuum Photomultiplier Tubes (PMT), or Hybrid Photon Detectors (HPD). The time resolution limit of current APD, PMT or HPD for single photo-electron detection is about 100 ps full width at half maximum (FWHM). The dead time limit of these devices is a few tens of ns.

Improvement of the temporal resolution of single photon detectors has an impact on a broad variety of applications [8]. For example, it can increase data rates and transmission distances for both classical [9] and quantum [10-12] optical communication systems, improve spatial resolution in laser ranging and enable observation of shorter-lived fluorophores in biomedical imaging [2].

In recent years, superconducting nanowire single-photon detectors (SNSPD) [13, 14] have provided single photon counting detectors with excellent temporal resolution below 15 ps [15-21].

However, timing systems based on Radio Frequency (RF) fields, the so-called RF timing technique, can provide a resolution of order 1 ps or better. The basic principle of the RF timing technique is the conversion of information in the time domain to a spatial domain by means of a high frequency RF field [22-26]. Streak Cameras, based on these principles, routinely operate in the ps and sub-ps time domain, but have substantial dead time associated with the readout system. In this paper we report a new type of RF timing technique, where the position sensor produces fast ns pulses, resulting in very small intrinsic dead time. The structure of the device, tests with thermo-electrons and further tests with a CW, UV photon beam are described in Section 2. In section 3 experimental results obtained with a RF synchronized femtosecond laser beam are presented. The time resolution and stability issues are discussed in Sections 4 and 5 respectively, followed by a summary and outlook.

## 2. A new RF timing technique of electrons and photons

The RF timing technique is schematically shown in Fig. 1. It is divided into several parts: UV light source, RF timing tube, RF source, position-sensing electronics, data acquisition system, power supplies and vacuum system. The crucial part is the RF timer whose components are mounted in a tube and operated at vacuum $10^{-6}$ Torr or less.

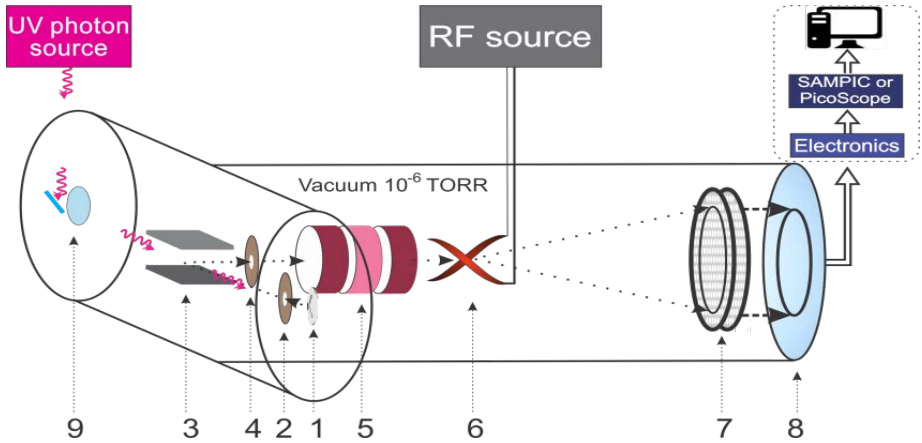

Figure 1: Schematic of the RF timing technique. 1-Tantalum disc cathode; 2-accelerating electrode; 3-permanent magnet; 4-collimator; 5-electrostatic lens; 6-helical RF deflector; 7-dual chevron-type MCP; 8-delay-line anode; 9-UV photon transmission window.

We have used the model ES-042 Tantalum Disc Cathode of Kimball Physics [30]. The diameter is 0.84 mm and the work function is 4.1 eV. The cathode, heated to 2200 K, emits a continuous low-energy electron beam with an energy spread of ~0.6 eV. The emitted electrons are accelerated by a voltage V ~2.5 kV applied between the cathode (1) and an accelerating electrode (2), which is a copper disc with a 1.5 mm diameter hole in the center, located about 5mm from the cathode. The accelerated electrons are deflected through 90 deg. by the permanent magnet (3) and pass through a collimator (4) with a diameter of ~1.5 mm, before entering the electrostatic lens (5). This focuses the electrons on the position-sensitive detector (PSD) consisting of a dual chevron MCP (7) and delay-line (DL) anode (8). However before reaching the PSD, the electrons pass through the RF deflection system (6) consisting of half period, helical electrodes [22] and a 500 MHz RF power source. The deflector has a diameter of 8 mm, a length of 30 mm and its exit point sits ~120 mm from the MCP system (7). The electrons are multiplied by a factor ~$10^6$ in the MCP system and the resulting electron cloud hits the DL anode (8), producing position-sensitive signals with rise times of a few ns. The active diameter of the MCP is 25mm. The helical-wire DLD40 DL anode [27] is a high-resolution 2D-imaging and timing device with a 40 mm active diameter. A photograph of the test experimental setup is presented in Fig. 2.

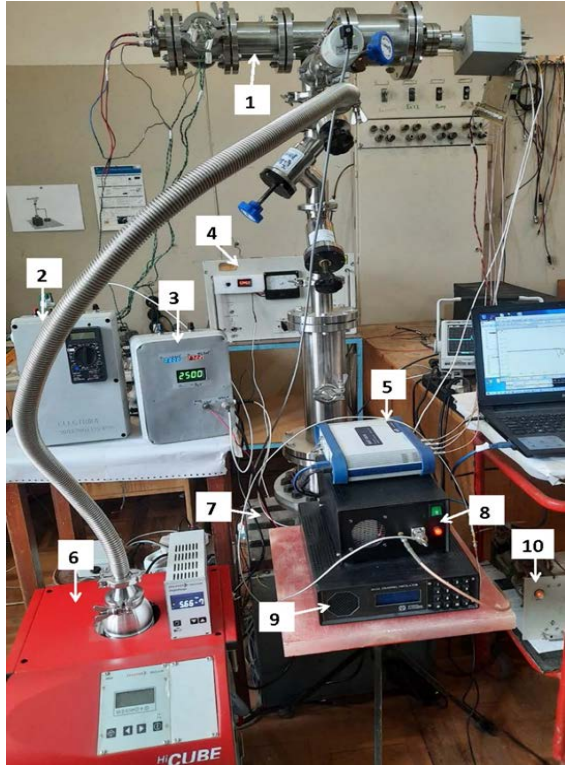

Figure 2: General view of the RF Timer and the test experimental setup. 1-RF Timer prototype; 2-power supply for electrostatic lens, 3-HV power supply for accelerating electrode and MCP detector; 4-HV power supply for "Titanium" vacuum pump; 5-PICOSCOPE; 6-Pfeiffer vacuum pump; 7-"Titanium" vacuum pump; 8-RF amplifier; 9-RF power source; 10-power supply for signal amplifiers. Internal components of the RF Timer displayed in Fig.1 are situated inside the high-vacuum prototype assembly (component 1) at the top of the photograph.

Hit coordinates are sensed by differential wire pairs, formed by a collection (signal) wire and a reference wire. The four wire pairs feed through the vacuum container for external processing. These signals are then amplified by custom-built fast amplifiers and sent to the Data Acquisition (DAQ) system consisting of digitizing oscilloscope SAMPIC [28] or PICOSCOPE [29] hardware and running software to reconstruct the electron hit position on the MCP. The position of the detected electrons is encoded in the difference in signal arrival time ($t_{x1}, t_{x2}$ and ($t_{y1}, t_{y2}$ for each end of each parallel-pair delay-line. Using the appropriate calibration coefficients $Q_x = 0.793$ mm/ns and $Q_y = 0.735$ mm/ns, the time (ns) is transformed to position (mm): $X = Q_x(t_{x1} - t_{x2})$ and $Y = Q_y(t_{y1} - t_{y2})$. A typical amplified signal from the DL anode, along with 2D images of the focused electron beam for RF power OFF and RF power ON are shown in Fig. 3.

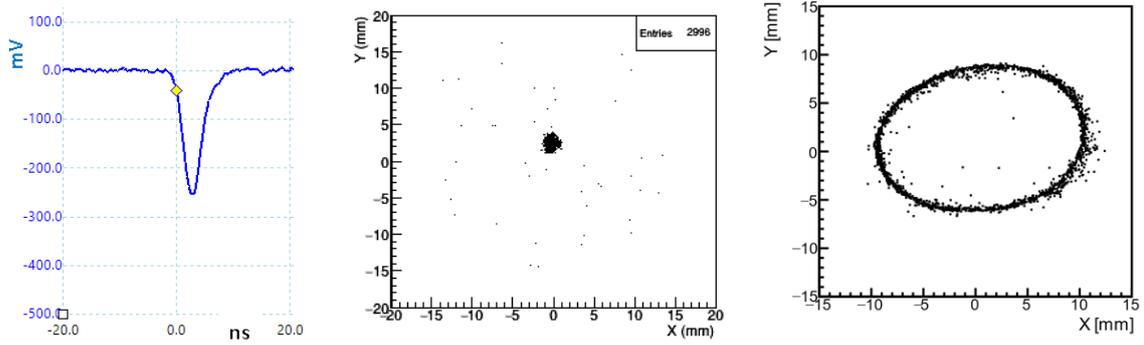

Figure 3: (Left) typical amplified signal from the DL anode; (middle) 2D image of the focused electrons (RF is OFF); (right) 2D image of the scanned electrons (RF is ON).

Pickup on the anode induced by the RF power is negligible and does not affect the reconstruction of the image (time) of the RF scanned electrons. The scanned ellipse of Fig. 3 confirms the previous work of Ref. [23] and shows that the present apparatus is operating correctly.

A possible application of this new timing technique is in the Radio Frequency Photo Multiplier Tube (RFPMT) [23]. The work function of the Tantalum cathode is 4.1 eV [30], so that photons with energies exceeding 4.1 eV (wavelength < 302 nm) can produce electrons. The RF timing tube has a quartz-glass window (9 in Fig. 1), which also allows testing of this new timing technique by using UV photons. The studies started with a UV diode, operating at 273 nm (4.54 eV) in CW mode. Photons from the diode transmitted through the quartz glass window, hit the cathode and the produced photoelectrons (PE) are transported through the RF timer as in test with thermo-electrons. Typical signals from the anode, and the obtained hit-position distributions, in cases where the RF source is OFF and ON, are very similar to those obtained with thermo-electrons (Fig. 3).

In the next step experimental studies continued at the CANDLE, AREAL laser facility [31]. The AREAL laser facility provided UV, 258 ± 1 nm (4.8 eV) photon bunches synchronized to a 500 MHz oscillator. The length of the photon pulses is adjustable from 0.45 to 10.0 ps (FWHM) and their rate lies in the range 1-100000 Hz. In our studies we used 0.45 ps duration photon pulses at a repetition rate of 100 Hz. The experiment started with the laser photon beam unsynchronized with the RF oscillator. The resulting hit-position distributions were very similar to those obtained with thermo-electrons (Fig.3) or the CW UV diode. It follows from the X and Y distributions presented in Fig. 4 (RF power off) that the position resolution is about 100 μm. In streak cameras this RF-off mode of operation is known as focus mode.

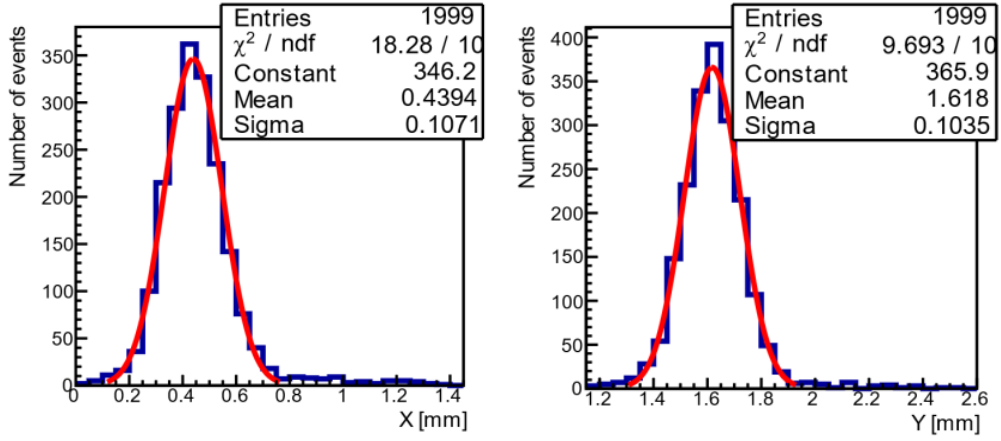

Figure 4: Distribution of X and Y coordinates of anode hit positions (RF is OFF), produced by the AREAL pulsed laser beam.

These results demonstrate the operational principles of the RFPMT. In such a photon detection system the sources of timing error are minimized because the produced PEs are timed before signal amplification and processing at the RFPMT output. The process of electron multiplication in this case is used to fix their coordinates on the scanning circle and consequently the corresponding phases of the applied RF field.

### 3. The RF timer in synchroscan mode

In the next step the RF timer was operated with the AREAL UV photon beam synchronized to the RF oscillator. The 0.45 ps duration photon pulses, phase locked to the 500 MHz oscillator at a repetition rate of 100 Hz, were directed to the Tantalum disc cathode and the 500 MHz RF was used to power the RF deflector. The RF signal $V_{RF}(t)$ can be written in the form:

$$V_{RF}(t) = V_{RF}^0 \sin[2\pi\nu_{RF} t + \varphi_{RF}(t) + \varphi_{RF}^0], \qquad (1)$$

where the amplitude $V_{RF}^0$ is assumed constant. The quantity $\nu_{RF} = n \times \nu^0$, where $\nu^0$ is the constant nominal frequency of the mode locked laser, $n$ is an integer and $t$ is the time referenced to the periodic laser pulse. The phase $\varphi_{RF}(t)$ contains the deviations, random and systematic, relative to the ideal oscillation $\nu^0$. In our case $\nu^0 = 50$ MHz (the frequency of the AREAL master oscillator) and $n = 10$ so that $\nu_{RF} = 500$ MHz. The quantity $\varphi_{RF}^0$ represents the nominal phase, which is constant for a given setup.

It is expected that the AREAL photon beam has the same time structure as the RF signal (eq.1) with the same random and systematic deviations, $\varphi_{RF}(t)$. Photoelectrons produced in the RFPMT cathode are accelerated, deflected by the magnet and focused on the PSD by the electrostatic lens. They pass through the RF deflector at time $t^i$ and fix the total phase of the RF oscillator somewhere on the scanning circle, for a given $\varphi_{RF}^0$:

$$\Phi_{RF}^i = 2\pi\nu_{RF}^0 t^i + \varphi_T(t^i) + \varphi_{RF}^0. \qquad (2)$$

Here the phase $\varphi_T(t^i)$ contains random and systematic deviations relative to the ideal oscillation, due to the RFPMT. The time $t^{i+1}$ for the next photon will be transposed to the phase

$$\Phi_{RF}^{i+1} = 2\pi\nu_{RF}^0 t^{i+1} + \varphi_T(t^{i+1}) + \varphi_{RF}^0. \tag{3}$$

For the ideal device, without any random or systematic drifts ($\varphi_T(t) = 0$), the produced PEs are perfectly in phase with the photon bunches. In this case, the sinusoidal RF signal from the AREAL master oscillator plays the role of time reference.

The timing jitter of the RFPMT may be measured from the PEs produced by the incident photon pulses of the laser. From the measured *X* and *Y* coordinates, the radius *R* and phase φ of the scanned electrons were determined by transforming to polar coordinates and converting φ to ns. For 500 MHz RF (T = 2 ns), 1° corresponds to 2ns/360° = 0.00555 ns. The 2D image of anode hit positions, as well as φ and *R* distributions, for a fixed phase of RF synchronization to the laser photon beam, are shown in Fig. 5. The black spot in the center of Fig. 5 (left) is a 2D image of the 2.5 keV electrons, obtained when the RF is OFF. The ellipse is an image of the scanned electrons, when the 500 MHz RF is ON, but not synchronized with the laser. The red spot on the ellipse corresponds to the phase distribution of RF-synchronized photoelectrons for a fixed phase. In this case, all photoelectrons have the same phase and, instead of a scanned circle, a spot on the circle is obtained. The spread in phase of these points represents the overall time resolution of the system, which includes factors related to the laser, the laser and RF oscillator synchronization device and the intrinsic time resolution of the RFPMT. It follows from the φ distribution presented in Fig. 5(middle) that the time resolution is about 10 ps.

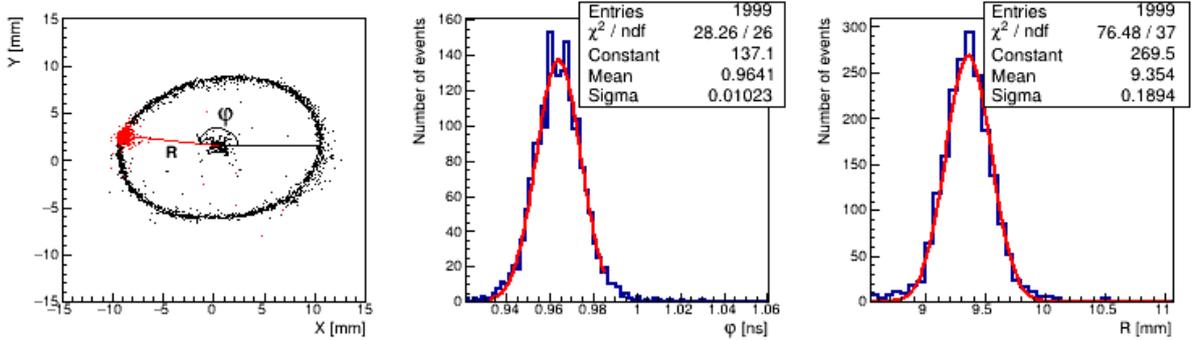

Figure 5: (Left) 2D image of anode hit positions; (middle) distribution of phase; (right) distribution of radius of the scanned electrons.

The final results with four fixed phases, located about 100 ps relative each other, are displayed in Fig. 6. The colored spots on the ellipse, labeled 1-4, correspond to phase distributions of RF-synchronized photoelectrons for the four different fixed phases. The distributions of these phases are shown in 1D in Fig. 6 (right). These results demonstrate the operational principles of the RFPMT [26], a new device for low-dead-time single photon counting, potentially with timing precision at the ps level.

Note that the helical deflection coil employed here was designed for a nominal RF of 500 MHz, but in production small deviations from the ideal shape gave circular scanning at a frequency of 490 MHz. At AREAL it was not possible to fine-tune the RF frequency and so at 500 MHz the scanning distribution is elliptical. The deviation from circular scanning is relatively small and does not affect the time resolution measurement of 10 ps significantly.

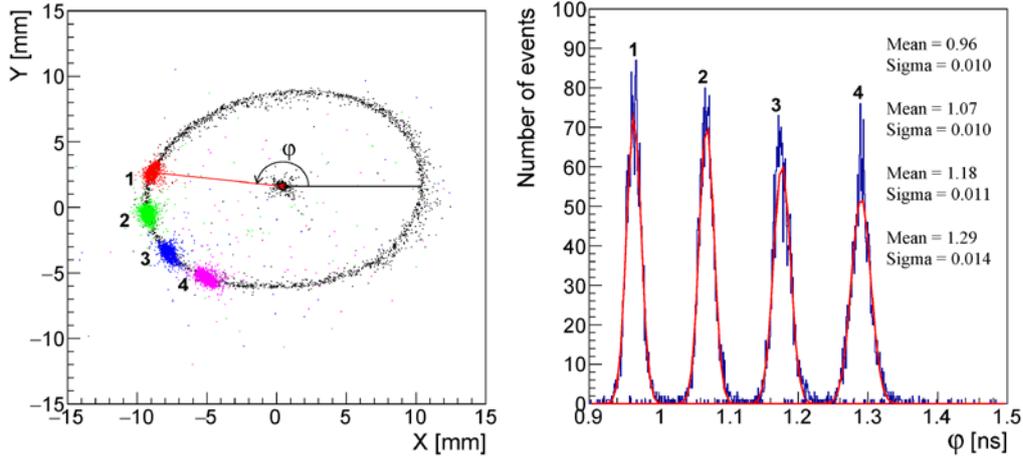

Figure 6: (Left) 2D images of anode hit positions; (right) phase distributions of the 500 MHz scanned synchronized photoelectrons. The different phases are labeled 1 – 4.

## 4. Factors contributing to the Time Resolution of the RFPMT

Several physical and technical factors contribute to the time resolution of the RFPMT in a synchroscan mode, which is expected to be similar to that of streak cameras [32]. Here we will discuss factors, that directly related to the RFPMT [33-36]. The first is related to the photoelectron emission and transport processes taking place in the photocathode. General quantum-mechanical considerations show that the inherent time scale of photoelectron emission should be much shorter than $10^{-14}$ sec. However, a time spread of the PEs $\Delta\tau_p$ can be caused by the finite thickness of the photocathode $\Delta l$. For typical photocathodes with $\Delta l = 10$ nm, $\Delta\tau_p \approx 10^{-13}$ sec. The second factor is the chromatic aberration due to the emitted PE's initial energy spread $\Delta\varepsilon$. The time resolution for $\Delta\varepsilon = 0.7$ ev, taking into account the parameters of our experimental setup, was simulated by means of the SIMION software package [37] and amounts to ~8 ps. The next factor is a technical time resolution of the circular RF deflector: $\Delta\tau_d = d/v$, where $d$ is determined by the size of the electron beam spot and the position resolution of the secondary electron detector. Parameter $v$ is the scanning speed ($v = 2\pi R/T$), where $T$ is the period of the RF field and $R$ is the radius of the circle of scanned electrons. In this case $T = 2$ ns, and $R$ is ~9.35 mm (Fig. 5 right). In effect $d$ is the dispersion of $R$ (0.19 mm), giving $\Delta\tau_d = 0.19/(2\pi \times 9.35) \times 2\text{ns} = 6.5 \times 10^{-12}$ sec. Therefore, the total internal time dispersion of the RFPMT is $(8^2 + 6.5^2)^{1/2}$ ps = 10.3 ps, which is in agreement with the measured resolution.

## 5. Time Stability

Time stability is a crucial factor for any high-precision timing system. The time stability of our timing system was investigated using the 500 MHz RF synchronized laser beam. A series of measurements at a fixed phase was carried out at ~10 min intervals. The mean values of the sequentially measured phase distributions is shown in Fig. 7, demonstrating that the time stability of the RFPMT over a period of ~1 hour is within the range of the statistical uncertainty which is about 0.5 ps, FWHM. Therefore, at the AREAL laser photon beam and with the RFPMT, a time correlated single photon counting can be realized with a resolution ≤10 ps.

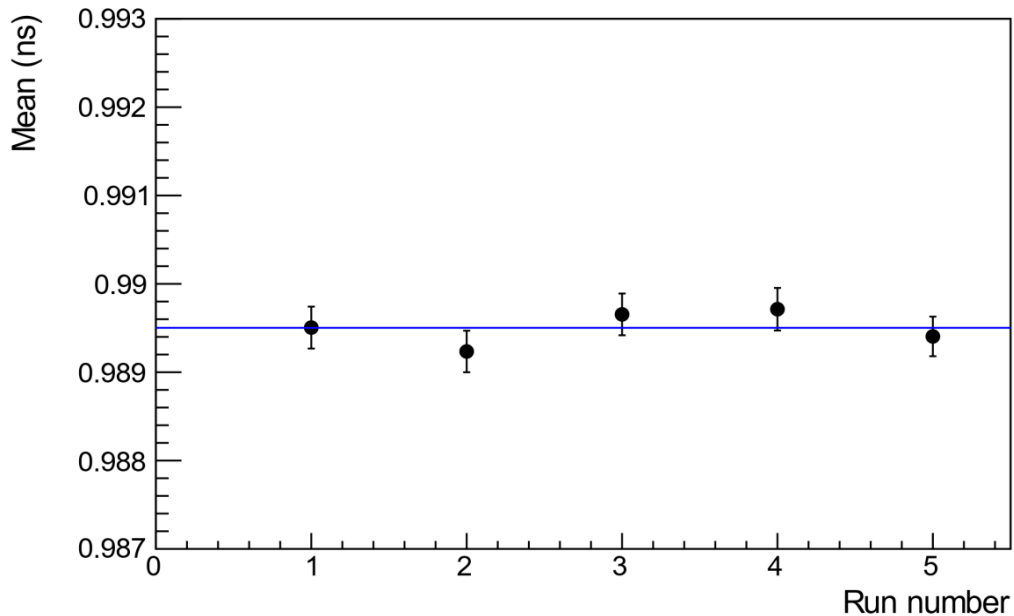

Figure 7: The mean values of sequentially measured time distributions in a one-hour period.

Since the data of Fig. 7 span a limited time period, an extended study over a longer time period with reduced statistical uncertainties is planned.

## 6. Summary and Outlook

This paper describes an RF timer of keV energy electrons. It is based on a helical deflector, which performs circular or elliptical sweeps of keV electrons by means of a 500 MHz radio frequency electromagnetic field. By converting the time of arrival of incident electrons to a hit position on a circle or ellipse, this device achieves extremely precise timing. Detection of the scanned electrons is implemented using a position sensitive detector based on microchannel plates and a delay line anode. Test studies started with thermo-electrons produced by a Tantalum disc cathode, which has a work function of 4.1 eV. This cathode was also used for experiments with 273 nm (4.54 eV) CW light diode and 258 nm (4.8 eV) femtosecond laser beams incident on the cathode to produce photo electrons. In this mode the RF timer became a radio frequency photomultiplier tube (RFPMT). The experiment with the RF synchronized femtosecond laser beam demonstrated ~10 ps time resolution for RFPMT. This 10 ps resolution is mainly due to the technical parameters of the prototype tube and RF deflector. With a view to achieving ps resolution or better we will continue to optimize these parameters, e.g. by, minimizing distance between the cathode and accelerating electrode, and increasing the frequency of the applied RF. Eventually a few hundred fs time resolution is expected for single photons with such an optimized device, operated with a 10 GHz RF.

Currently the intrinsic dead time of the RFPMT is determined by the propagation time of signals in a DL anode, which are around 20 ns in duration. This is already vastly better than any streak camera, but could be reduced to the ps level by the use of a fast pixelated detector.

Multiple single-photon events, spaced by a little as ~1 ps, could be detected during a single scan cycle by devices such as Timepix4 [38] which has 256×256 pixels of size is 55×55 μm. Developments by our collaboration of Timepix readout are already underway. Circular scanning has a limited time range, the inverse of the RF setting. We are investigating a dual-coil system, each coil operated at slightly different frequencies, $\omega_1$, $\omega_2$, where the resulting "beat" in deflection amplitude produces a spiral pattern rather than a circle. The time range to complete a spiral cycle (the inverse of the beat frequency) would be 100 ns for $\omega_1$=495 MHz and $\omega_2$=505 MHz.

After further development, the RFPMT has potential applications in many fields of science and industry, which include fundamental physics [39-41], high energy [42-44] and nuclear physics [45-49], chemistry, medical and biomedical imaging and material science [50,51]. Technological and industrial applications include accelerator physics [52], laser ranging, dynamic testing of integrated circuits and high-speed optical components for data storage and fiber optic telecommunication.

## 7. Acknowledgements

This work was partially supported by the International Science and Technology Center (ISTC) in the framework of scientific project A-2390, the State Committee of Science of the Republic of Armenia (Grants: 21T-2J133, 20TTCG-1C011,18Ap_2b05 and 21APP-2B012), the ARPA Institute, the ANSEF (Grant hepex-4954), the UK Science and Technology Facilities Council (Grants: ST/V00106X/1, ST/S00467X/1), the JSPS KAKENHI (Grants: 18H05459, 17H01121), the European Union's Horizon 2020 research and innovation program (Grant agreement No 824093). The Yerevan group thanks Prof. Ani Aprahamian, Dr. Hakob Panosian, Prof. Ara Apkarian, Dr. Arpine Piloyan, Dr. Ottmar Jagutzki, Dr. Dominique Breton, Dr. Viatcheslav Sharyy, Dr. Dominique Yvon, Dr. Narine Sarvazyan and staff of the AREAL laser facility for extremely helpful discussions and assistance.

## 8. Author Contribution

Amur Margaryan: supervision, conceptualization, methodology, resources, original draft writing. Vanik Kakoyan: management, conceptualization, investigation, software, resources. Simon Zhamkochyan: software, investigation, resources. Sergey Abrahamyan: software, investigation. Hayk Elbakyan: investigation, resources. Samvel Mayilyan: hardware. Henrik Vardanyan: electronics. Hamlet Zohrabyan: software. Lekdar Gevorgian: theory. Robert Ayvazyan: hardware. Artashes Papyan: electronics. Garnik Ayvazyan: hardware. Bagrat Grigoryan: validation. John Annand: conceptualization, investigation, writing-review and editing, resources. Kenneth Livingston: conceptualization, investigation, review and editing, resources. Rachel Montgomery: conceptualization, investigation, review and editing, resources. Patrick Achenbach: conceptualization, review and editing, resources. Josef Pochodzalla: conceptualization, review and editing, resources. Dimiter L. Balabanski: conceptualization, resources, Satoshi N. Nakamura: conceptualization, review and editing, resources.